
\input harvmac
\noblackbox
\newcount\figno
\figno=0
\def\fig#1#2#3{
\par\begingroup\parindent=0pt\leftskip=1cm\rightskip=1cm\parindent=0pt
\baselineskip=11pt
\global\advance\figno by 1
\midinsert
\epsfxsize=#3
\centerline{\epsfbox{#2}}
\vskip 12pt
\centerline{{\bf Figure \the\figno} #1}\par
\endinsert\endgroup\par}
\def\figlabel#1{\xdef#1{\the\figno}}
\def\pano{\par\noindent}
\def\smno{\smallskip\noindent}
\def\meno{\medskip\noindent}
\def\bigno{\bigskip\noindent}
\font\cmss=cmss10
\font\cmsss=cmss10 at 7pt
\def\rlx{\relax\leavevmode}
\def\inbar{\vrule height1.5ex width.4pt depth0pt}
\def\IC{\relax\,\hbox{$\inbar\kern-.3em{\rm C}$}}
\def\IR{\relax{\rm I\kern-.18em R}}
\def\IN{\relax{\rm I\kern-.18em N}}
\def\IP{\relax{\rm I\kern-.18em P}}
\def\ZZ{\rlx\leavevmode\ifmmode\mathchoice{\hbox{\cmss Z\kern-.4em Z}}
 {\hbox{\cmss Z\kern-.4em Z}}{\lower.9pt\hbox{\cmsss Z\kern-.36em Z}}
 {\lower1.2pt\hbox{\cmsss Z\kern-.36em Z}}\else{\cmss Z\kern-.4em Z}\fi}
\def\narrowplus{\kern -.04truein + \kern -.03truein}
\def\narrowminus{- \kern -.04truein}
\def\narrowminussub{\kern -.02truein - \kern -.01truein}
\def\a{\alpha}

\def\cl{\centerline}

\def\o#1{\overline{#1}}

\def\la{\langle}
\def\ra{\rangle}

\def\type{type$\ {\rm II}$}
\def\typeo{type$\ {\rm 0B}$}
\def\typeoa{type$\ {\rm 0A}$}
\def\typep{type$\ {\rm 0}$}
\def\typepp{type$\ {\rm 0}'$}
\def\Type{Type$\ {\rm I}\ $}
\def\typea{type$\ {\rm IIA}$}
\def\typeb{type$\ {\rm IIB}$}
\def\Typeb{Type$\ {\rm IIB}\ $}

\def\sqr#1#2{{\vcenter{\vbox{\hrule height.#2pt
 \hbox{\vrule width.#2pt height#1pt \kern#1pt
 \vrule width.#2pt}\hrule height.#2pt}}}}


\def\drawbox#1#2{\hrule height#2pt 
        \hbox{\vrule width#2pt height#1pt \kern#1pt 
              \vrule width#2pt}
              \hrule height#2pt}

\def\Fund#1#2{\vcenter{\vbox{\drawbox{#1}{#2}}}}
\def\Asym#1#2{\vcenter{\vbox{\drawbox{#1}{#2}
              \kern-#2pt       
              \drawbox{#1}{#2}}}}
 
\def\funda{\Fund{6.5}{0.4}}
\def\asymm{\Asym{6.5}{0.4}}

\def\basymm{\overline{\asymm}}


\lref\rkletsy{I.R. Klebanov and A.A. Tseytlin, {\it D-Branes and Dual Gaue 
theories in Type 0 Strings}, hep-th/9811035.} 

\lref\covlat{W. Lerche, D. L\"ust and A.N. Schellekens,
{\it Chiral Four-dimensional Heterotic Strings from Self-dual Lattices},
Nucl.\ Phys.\ {\bf B287} (1987) 477.} 

\lref\rbergab{O. Bergman and M.R. Gaberdiel, {\it A Non-Supersymmetric Open 
String Theory and S-Duality}, hep-th/9701137.}

\lref\rangel{C. Angelantonj, {\it Non-Tachyonic Open Descendants of the 
0B String Theory}, hep-th/9810214.}

\lref\rsagn{A. Sagnotti, {\it Some Properties of Open String Theories},
 hep-th/9509080 \semi
A. Sagnotti, {\it Surprises in Open String Perturbation Theory},
hep-th/9702093.}

\lref\rsagbi{A. Sagnotti, M. Bianchi,
{\it On the Systematics of Open String Theories},
Phys.\ Lett.\ {\bf B247} (1990) 517}

\lref\rsagb{A. Sagnotti, M. Bianchi,
{\it Twist Symmetry and Open String Wilson Lines},
Nucl.\ Phys.\ {\bf B361} (1991) 519.} 

\lref\rgimpol{
E.G.\ Gimon and J.\ Polchinski, {\it Consistency Conditions
for Orientifolds and D-Manifolds}, Phys.\ Rev.\ {\bf D54} (1996) 1667,
hep-th/9601038.}

\lref\rgimjo{ E.G.\ Gimon and C.V.\ Johnson, {\it K3 Orientifolds},
Nucl.\ Phys.\ {\bf B477} (1996) 715, hep-th/9604129 \semi
A. Dabholkar, J. Park, {\it Strings on Orientifolds},
Nucl.Phys. {\bf B477} (1996) 701, hep-th/9604178.}

\lref\rfisus{W.\ Fischler and L.\ Susskind, 
{\it Dilaton Tadpoles, String Condensates, and Scale Invariance}, 
Phys.\ Lett.\ {\bf B171} (1986) 383;  
{\it Dilaton Tadpoles, String Condensates, and Scale Invariance II}, 
Phys.\ Lett.\ {\bf B173} (1986) 262.}  

\lref\rseibwit{L. Dixon and J. Harvey, {\it String Theories in Ten Dimensions
Without Space-Time Supersymmetry}, Nucl.\ Phys.\ {\bf B274} (1986) 93 \semi
N. Seiberg and E. Witten, {\it Spin Structures in String Theory}, 
Nucl.\ Phys.\ {\bf B276} (1986) 272. }

\lref\rbackb{  M. Bianchi, G. Pradisi and A. Sagnotti, 
{\it Toroidal Compactification and Symmetry Breaking in Open-String 
    Theories}, 
    Nucl.Phys. {\bf B376} (1992) 365 \semi
Z. Kakushadze, G. Shiu and S.-H.H Tye, {\it \Typeb Orientifolds
 with NS-NS Antisymmetric Tensor Backgrounds}, 
Phys. Rev. {\bf D58} (1998) 086001, hep-th/9803141.}

\lref\rwitti{E. Witten, {\it Small Instantons in String Theory},
Nucl.\ Phys.\ {\bf B460} (1996) 541, hep-th/9511030. }

\lref\rkaku{ Z. Kakushadze, G. Shiu, S.-H. H. Tye, {\it
   Type IIB Orientifolds, F-theory, Type I Strings on Orbifolds and 
    Type I - Heterotic Duality}, Nucl. Phys. {\bf B533} (1998) 25,
    hep-th/9804092. \semi
   Z. Kakushadze, {\it Non-perturbative Orientifolds}, hep-th/9904007.}

\lref\rafiv{G. Aldazabal, A. Font, L.E. Ib\'a\~nez, G. Violero, {\it
D=4, N=1, Type IIB Orientifolds}, Nucl. Phys. {\bf 536} (1998) 29, 
hep-th/9804026.}

\Title{\vbox{\hbox{hep--th/9904069}
 \hbox{HUB--EP--99/17}}}
{\vbox{ \hbox{Tachyon-free Orientifolds of Type 0B} 
\vskip 0.4cm
\hbox{\phantom{eee} Strings in Various Dimensions}}}
\centerline{Ralph Blumenhagen${}^1$, Anamar\'{\i}a Font${}^2$ and 
Dieter L\"ust${}^3$}
\bigskip
\centerline{\it ${}^{1,3}$ Humboldt-Universit\"at Berlin, Institut f\"ur 
Physik,}
\centerline{\it  Invalidenstrasse 110, 10115 Berlin, Germany }
\smallskip
\centerline{\it ${}^2$ Departamento de F\'{\i}sica, Facultad de Ciencias,
 Universidad Central de Venezuela,}
\centerline{\it A.P. 20513, Caracas 1020-A, Venezuela }
\centerline{\it and}
\centerline{\it Centro de Astrof\'{\i}sica Te\'orica, Facultad de Ciencias,}
\centerline{\it Universidad de Los Andes, Venezuela} 
\smallskip
\bigskip
\bigskip
\centerline{\bf Abstract}
\noindent
We construct non-tachyonic, non-supersymmetric orientifolds 
of \typeo\ strings in 
ten, six and four space-time dimensions. Typically, these models have  
unitary gauge groups with charged massless fermionic and bosonic matter 
fields. However, generically there remains an uncancelled dilaton tadpole.

\footnote{}
{\pano
${}^1$ e--mail:\ blumenha@physik.hu-berlin.de
\pano
${}^1$ e--mail:\ afont@fisica.ciens.ucv.ve
\pano
${}^3$ e--mail:\ luest@physik.hu-berlin.de
\pano}
\Date{04/99}
\newsec{Introduction}

Due to its possible application to the dynamics of non-supersymmetric 
gauge theories in four dimensions, \typeo\ models have received much
attention during the last months. 
Unfortunately, these models are plagued by the appearance of a tachyonic
mode in the closed string sector. However the effectice gauge theory on
the D-branes of \typeo\ models \rkletsy\ is nevertheless tachyonfree and hence
consistent.
The couplings of the tachyon in the 
effective low
energy field theory allow the dilaton to depend on the distance to
the branes in just the right way to be consistent with the one-loop running
of the gauge coupling in non-supersymmetric gauge theories. 

In order to study consistent non-supersymmetric string vacua, 
one should of course get
rid of all tachyonic modes. 
Tachyon-free, non-supersymmetric heterotic strings in four dimensions
with chiral, massless fermions were already constructed some time ago
for example in the covariant lattice formalism \covlat.
In the context of the \typeo\ string 
one should include some projections modding out the tachyon. 
In \refs{\rsagbi, \rbergab}
the standard orientifold
of \typeo\ was considered leading to a model containing  still the
tachyon in the closed string sector as well as new tachyonic modes 
coming from open strings stretched between D9 and $\o{\rm D9}$ branes.
The latter objects needed to be introduced into the background to satisfy
tadpole cancellation conditions. In \rbergab\  the D-brane content of 
\typeo\ and its orientifold \typep\ was
derived using the boundary state formalism. From the partition function
of \typeo\ it is clear that all Ramond-Ramond fields are doubled with respect
to \typeb\ leading to a doubling of the D-branes as well.      

In \rsagn\ it was observed on the level of conformal field theory partition
functions that there exists another Klein-bottle projection which leads
to a non-tachyonic ``Type I descendant'' called \typepp\ in the following.
This model has a non-zero dilaton tadpole which however 
can be cancelled by the Fischler-Susskind \rfisus\ mechanism.
Compactifications of \typepp\ to six and four space-time dimensions were 
considered in \rangel. However, the microscopic description in particular 
the D-brane contents of these models is obscure in \refs{\rsagn, \rangel}. 

In this paper we revisit the non-tachyonic orientifold of \typeo,
clarify the origin of a second orientifold projection in terms of a
different world-sheet parity $\Omega'$, and determine
the D-brane contents of this model.  
Moreover, we consider \typeo\ orientifolds on
$T^{10-d}/\{G_1,\Omega'G_2\}$, $d=6,4$, and derive the anomaly
free massless spectra in specific models.
All of these models have in common some phenomenological appealing features
like a pure bosonic always unitary gauge sector with both bosonic 
and {\it fermionic} charged matter. As in the parent \typeo\ model
the gravity sector is purely bosonic. 

Moreover, we study the effective theory on the D3-branes in more detail 
finding
some pecularities reflecting the fact that D9 and D3-branes
repel each other. 

\newsec{Review of  \typeo\ model}

In the construction of critical superstring theories one has to implement  the
GSO projection 
\eqn\proja{ P_{NSNS}={1\over 4}(1+(-1)^{F_L})(1+(-1)^{F_R}), \quad\quad
            P_{RR}={1\over 4}(1+(-1)^{F_L})(1\mp (-1)^{F_R}). }
The two 
possible choices for the GSO projection in the Ramond sector  lead to
\typea\ and \typeb, respectively. However, it is known since 1986 \rseibwit\ 
that using only the projection 
\eqn\projb{ \o{P}_{NSNS}={1\over 2}(1+(-1)^{F_L+F_R}), \quad\quad
            \o{P}_{RR}={1\over 2}(1\pm(-1)^{F_L+F_R}) }
still leads to modular invariant partition functions on the torus
\eqn\modinv{ Z_T={1\over 2}{ |f_3|^{16} + |f_4|^{16} + |f_2|^{16} 
                \over |f_1|^{16} }, }
called \typeo\ and \typeoa\ depending on the sign in $\o{P}_{RR}$.
These ten dimensional string theories are pure bosonic hence 
non-supersymmetric and contain a tachyon in the spectrum.
The non-supersymmetric model \modinv\  can also be considered as the 
orbifold of \type\ by the 
space-time fermion operator $(-1)^{F_S}$, where the tachyon makes 
its appearance in the twisted sector. 
At the massless level the NS-NS sector contributes a  graviton $G_{\mu\nu}$,
a dilaton $\Phi$ and an antisymmetric 2-form $B_{\mu\nu}$ to the 
spectrum. For \typeo\ the RR sector contributes two further scalars 
$\Phi^{1,2}$, two antisymmetric 2-forms 
$B^{1,2}_{\mu\nu}$ and a 4-form $D_{\mu\nu\rho\sigma}$. 
In \typeoa\ models there are 
two 1-forms $A_\mu^{1,2}$ and two antisymmetric 3-forms 
$C^{1,2}_{\mu\nu\rho}$ in the RR sector. 
Since compared to the \type\ models all fields in the RR sector are doubled,
one expects the D-brane content to be doubled, as well. The explicit 
form of the corresponding boundary states was derived in \rbergab, which we
will briefly review for later reference.   \pano
The boundary state of a D$p$-brane in \type\ is a sum of four terms
\eqn\bound{ |{\rm D}p\ra = {1\over 2}\left( | {\rm D}p,\eta=+1\ra_{NSNS} - 
              | {\rm D}p,\eta=-1\ra_{NSNS} +
         | {\rm D}p,\eta=+1\ra_{RR}  + | {\rm D}p,\eta=-1\ra_{RR}\right) }
and is invariant under the GSO projections of \typea\ for even p and
of \typeb\ for odd p. 
The individual terms are given by
\eqn\boundb{\eqalign{  | {\rm D}p,\eta\ra_{NSNS}=
       \int (\prod_{\nu=p+1}^9 dk^\nu)\ &{\rm exp}\left\{ \sum_{n=1}^\infty
              \left[  -{1\over n}\sum_{\mu=2}^{p}
                  \alpha^\mu_{-n} \tilde\alpha^\mu_{-n} + {1\over n}
                 \sum_{\mu=p+1}^{9}
                  \alpha^\mu_{-n} \tilde\alpha^\mu_{-n}
                \right] \right\} \cr
               &{\rm exp}\left\{ i\eta \sum_{r>0}
              \left[  -\sum_{\mu=2}^{p}
                  \psi^\mu_{-r} \tilde\psi^\mu_{-r}  
                + \sum_{\mu=p+1}^{9}
                  \psi^\mu_{-r} \tilde\psi^\mu_{-r}
                \right] \right\}|\vec{k},\eta\ra_{NSNS} \cr   }}             
and analogously for the RR sector, where one has to define the RR vacuum in
a consistent way.
However, in the \typeoa\  and \typeo\ theories each of the four terms 
in \bound\ is invariant
by itself under the $\o{P}$ projection, leading
to four possible boundary states for each $p$ 
\eqn\boundc{|{\rm D}p,\eta,\eta'\ra=  
                     | {\rm D}p,\eta\ra_{NSNS} + 
                      | {\rm D}p,\eta'\ra_{RR} }
and their corresponding anti-branes for which  the sign in front of the
Ramond-Ramond  boundary state is reversed. 
The action of the world-sheet fermion number operator $(-1)^{F_R}$ 
exchanges the $|{\rm D}p,\eta,\eta'\ra$ brane with the 
$|{\rm D}p,-\eta,-\eta'\ra$ brane.
Using the boundary states \boundc\  one first computes the cylinder amplitude
in tree channel 
\eqn\cyl{ \tilde{A}={1\over 2}\int_0^\infty {dl\over 2}\ 
     \la {\rm D}p,\eta_1,\eta_1'| 
         \,  e^{-l H_{cl}}\, | {\rm D}p,\eta_2,\eta_2'\ra} 
and then transforms to loop channel 
\eqn\cyla{ A_{(\eta_1,\eta_1'),(\eta_2,\eta_2')}= 
\int_0^\infty \, {{dt}\over t} \, 
                 {\rm Tr}_{open} \left[ \textstyle{{(1+(-1)^F)\over 2}\, 
                 {(1+(-1)^{F_S})\over 2}}\  e^{-2\pi t L_0} \right]. }
For the amplitudes we will be interested in this yields the following
result
\eqn\cylb{  A_{(\eta_1,\eta_1'),(\eta_1,\eta_1')}=
{V_{p+1}\over {(8\pi^2 \a')^{{p+1}\over 2} }} \, 
       \int_0^\infty {dt\over t^{p+3\over 2}}\
    { f_3^8(e^{-\pi t}) - f_4^8(e^{-\pi t}) \over f_1^8(e^{-\pi t})} }
for open strings stretched between the same type of D$p$-branes. 
However, for open strings stretched between opposite D$p$-branes one obtains
\eqn\cylbb{ A_{(\eta_1,\eta_1'),(-\eta_1,-\eta_1')}=
      {V_{p+1}\over {(8\pi^2 \a')^{{p+1}\over 2} }} \, 
    \int_0^\infty {dt\over t^{p+3\over 2}}\
    {  -f_2^8(e^{-\pi t}) \over f_1^8(e^{-\pi t})}, }
where the extra minus sign appears due to the action 
\eqn\actio{ (-1)^{F_R}| {\rm D}p,\eta\ra_{NSNS}=-|{\rm D}p,-\eta\ra_{NSNS} .}
$V_{p+1}$ is the regularized D$p$-brane volume.
From \cylb\ and \cylbb\ one derives that for open strings stretched between 
the same D$p$-branes the world-sheet fermions have half-integer
mode expansion whereas for open strings stretched between a 
$|{\rm D}p,\eta,\eta'\ra$ 
and a $|{\rm D}p,-\eta,-\eta'\ra$ brane they have integer mode expansion. 
In the latter case fermionic zero modes appear leading
to space-time fermions. \pano
In \rbergab, and originally in \rsagn, an orientifold of \typeo\ was studied,
with the world-sheet parity operation acting in the closed string sector as
follows
\eqn\omegap{\eqalign{ &\Omega \alpha_n  \Omega = \tilde\alpha_n ,\cr
                     &\Omega \psi_r  \Omega = \tilde\psi_r ,\cr
                     &\Omega \tilde \psi_r  \Omega = -\psi_r ,\cr
    &\Omega |0\ra_{NSNS}  = |0\ra_{NSNS}  .}}
Thus, the tachyon survived the projection and the Klein-bottle only produced 
a NS-NS tadpole which was partially cancelled by introducing the same number
of $|D9,+,+\ra$ and
anti- $|D9,+,+\ra$ branes in the background.   
It was not possible to cancel the tachyon tadpole. 
{}From open strings stretched between the different 9-branes they got
gauge bosons of $SO(32)\times SO(32)$ and a further tachyon transforming
in the bi-fundamental $({\bf 32},{\bf 32})$ representation of the gauge group.
\pano
Moreover, in \rbergab\ it was shown that D1, D5 and D9-branes do survive the 
orientifold
projection and that the world-volume theory on the D1-branes is in agreement
with the world-sheet theory of the bosonic string compactified on an $SO(32)$
lattice to ten dimensions. The latter observation led to the conjecture that
the two models might provide a string-weak dual pair. 
We will now show  in section 3  that there exists a different
orientifold projection leading to the tachyon-free model first discussed
in \rsagn. 

\newsec{The \typepp\  orientifold}

Instead of \omegap\ one can define the action of $\Omega$ in the closed string
sector as follows
\eqn\omegapp{\eqalign{ &\Omega' \alpha_n  \Omega' = \tilde\alpha_n ,\cr
                     &\Omega' \psi_r  \Omega' = \tilde\psi_r ,\cr
                     &\Omega' \tilde \psi_r  \Omega' = \psi_r ,\cr
    &\Omega' |0\ra_{NSNS} = -|0\ra_{NSNS} ,}}
which apparently means $\Omega'=\Omega\, (-1)^{F_R}$. In the \typeb\ theory
this makes no difference for the action of $\Omega'$ and $\Omega$ on physical
states is identical. On the contrary, in the non-supersymmetric \typeo\ theory
the two actions define two completely different models as can be seen
from the action of $\Omega'$ on the NS vacuum, which projects
out the tachyon. 

\subsec{The Klein bottle amplitude}

In doing an orientifold there generically appear tadpoles which ought to
be cancelled by contributions from an open string sector.  
Therefore, we compute the Klein bottle amplitude for this model
\eqn\klein{\eqalign{ K&= 32\, c \, \int_0^\infty {dt\over t^6}\, 
                  {\rm Tr} \left[
                  \Omega'\, \textstyle {{1+(-1)^{F_L+F_R} \over 2}}\, 
                 (-1)^{F_S}\, 
                   e^{-2\pi t(L_0+\o{L}_0)} \right] \cr
                  &\cr
         &= - 32 \, c \,\int_0^\infty {dt\over t^6}\, {f_4^8(e^{-2\pi t})
                      \over f_1^8(e^{-2\pi t}) } ,}}
where $c = V_{10}/(8\pi^2 \a')^5$.
The transformation into tree-channel  leads to
\eqn\kleinb{ \widetilde{K}=- 2^{11} \, c \ \int_0^\infty dl\, 
                    {f_2^8(e^{-2\pi l})
                      \over f_1^8(e^{-2\pi l})  }}
showing that there is only a contribution from RR 10-form exchange. 
The appropriate branes one has to introduce to cancel this tadpole
must be charged under the RR 10-form, which 
are of course D9-branes. However, due to the action of the right handed
world-sheet fermion number operator $(-1)^{F_R}$ on D9-branes we are
forced to introduce the same number N of  
$|Dp,\eta,\eta'\ra$ and $|Dp,-\eta,-\eta'\ra$ branes. Without loss of
generality, we choose $9=|D9,+,+\ra$ and $9'=|D9,-,-\ra$ branes in 
the following.  The next step is to compute the cylinder and M\"obius 
strip amplitudes for these branes. 

\subsec{The Cylinder amplitude}

The annulus amplitude is defined as
\eqn\anna{ A=c\, \int_0^\infty {dt\over t^6}\, {\rm Tr}_{99,9'9',99',9'9}
                 \left[ \textstyle{{1+(-1)^{F} \over 2} {1+(-1)^{F_S}\over 2}}
                   e^{-2\pi t L_0} \right] }
where the space time fermion number projection implies that only
the open string NS sector contributes.
Using the results \cylb\ and \cylbb\ it is straightforward to determine
the individual contributions  
\eqn\annb{\eqalign{ &A_{99}=A_{9'9'}=c\, \int_0^\infty {dt\over t^6}\, 
           {N^2\over 2}\,{f_3^8(e^{-\pi t})-f_4^8(e^{-\pi t})
                      \over f_1^8(e^{-\pi t}) } ,\cr
             &A_{99'}=A_{9'9}=-c\, \int_0^\infty {dt\over t^6}\, 
           {N^2\over 2}\,{f_2^8(e^{-\pi t})
                      \over f_1^8(e^{-\pi t}) } \cr }}
adding up to the complete tree channel cylinder amplitude
\eqn\annc{\widetilde{A}=c\, \int_0^\infty {dl}\, 
           2\,N^2\,{f_3^8(e^{-2\pi l})-f_4^8(e^{-2\pi l})-f_2^8(e^{-2\pi l})
                      \over f_1^8(e^{-2\pi l}) }  .}
Up to a factor of two this is identical to the cylinder amplitude
of two D9 branes of \typeb. Therefore, even without supersymmetry the
overall force between 
the same number of D9 and D$9'$-branes vanishes, hence  permitting  a stable 
configuration of these branes. Note, that the tree channel amplitude \annc\ 
contains a dilaton tadpole as well as a RR 10-form tadpole. 

\subsec{The M\"obius amplitude}

Finally, we have to compute the M\"obius strip amplitude 
\eqn\moeba{ M=c\, \int_0^\infty {dt\over t^6}\, {\rm Tr}_{99,9'9',99',9'9}
                 \left[ \Omega'\, \textstyle{{1+(-1)^{F} \over 2} 
                   {1+(-1)^{F_S}\over 2}}
                   e^{-2\pi t L_0} \right] ,}
where again there is no contribution from the open string R sector. 
Since $\Omega'$ exchanges D9 and D$9'$-branes, the only non-zero
contribution in \moeba\ is from open strings stretched between
D9 and D$9'$-branes. 
For the loop channel M\"obius amplitude one gets
\eqn\moebb{ M=c\, \int_0^\infty {dt\over t^6}\, {1\over 2}\, 
                { f_2^8(i e^{-\pi t})
                  \over f_1^8(i e^{-\pi t}) }\, {\rm Tr}(\gamma_{\Omega'}^T
                 \gamma_{\Omega'}^{-1}),}
where $\gamma_{\Omega'}$ describes the action of $\Omega'$ on the Chan-Paton
factors. 
Since the resulting tree channel  amplitude is given by
\eqn\moebc{\widetilde{M}= c\, \int_0^\infty {dl}\, 2^6 \, 
               { f_2^8(i e^{-2\pi l})
                    \over f_1^8(i e^{-2\pi l}) }\, {\rm Tr}(\gamma_{\Omega'}^T
                      \gamma_{\Omega'}^{-1})  , }
there is only a contribution to the RR exchange tadpole.

\subsec{Tadpole cancellation}
 
Adding up all the three contributions, the tadpole cancellation 
condition for the RR 10-form potential reads
\eqn\tada{ N^2 - 32\, {\rm Tr}(\gamma_{\Omega'}^T \gamma_{\Omega'}^{-1}) +
32^2 =0.}
However, from  NS-NS exchange we are left with an uncancelled
dilaton tadpole
\eqn\tadb{ c\, 
      \int_0^\infty dl\, 2\, N^2\,{f_3^8(e^{-2\pi l})-f_4^8(e^{-2\pi l})
                      \over f_1^8(e^{-2\pi l}) } =
      \int_0^\infty dl\ \left( 32\, c\, N^2 + O(e^{-2\pi l})\right)  . }
However, it was shown in \rfisus\
that a NS-NS tadpole does not render the theory inconsistent for it means
that we are not expanding around the true vacuum. One has to introduce
a non-constant dilaton and a tree level cosmological constant to cancel
such a tadpole. 
In  section 4 we will mention that such a tadpole can possibly also 
be cancelled
by introducing lower dimensional D-branes in the background.\pano
The RR tadpole cancellation condition \tada\ implies that the
matrix $\gamma_{\Omega'}$ has to be  symmetric. We are free to choose
\eqn\tadc{  \gamma_{\Omega'}=\left(\matrix{  0  &  1  \cr
                                             1  & 0 \cr} \right)_{2N,2N}, }
where the first $N$ entries denote  the D9-branes and the second $N$ entries
denote the D$9'$-branes. The off-diagonal choice in \tadc\ reflects the fact 
that $\Omega'$ exchanges D9 and D$9'$-branes.  
Plugging \tadc\ into \tada\ one realises that  the RR tadpole
is cancelled for precisely 32 D9 and 32 D$9'$-branes.

\subsec{Massless spectrum}
At the massless level in the NS-NS closed string sector  $\Omega'$ acts
in the same way as $\Omega$ leaving the graviton and the dilaton invariant,
whereas the antisymmetric 2-form is projected out. 
In the R-R sector before the projection one had
\eqn\massa{\eqalign{ C^8\otimes \tilde{C}^8 &= 1 + 28 +35^+ ,\cr
                     S^8\otimes \tilde{S}^8 &= 1 + 28 +35^-  .}}
Since $C^8\otimes \tilde{C}^8$ is invariant under $(-1)^{F_R}$,  
these states have to
be antisymmetrized under $\Omega'$ giving one 2-form field. 
However,  $(-1)^{F_R}$ gives an extra minus sign when acting on
$S^8\otimes \tilde{S}^8$, so that these states are symmetrized leading to
a scalar and an anti-selfdual 4-form  potential. 
Thus, the closed string spectrum is the bosonic part of \typeb, in particular
one expects the brane contents to be the same as in \typeb. 
\pano
In the open string sector, strings stretched between two D9 or
two D$9'$-branes, respectively, carry the massless mode
\eqn\massb{  \psi^\mu_{-{1\over 2}}|0\ra_{NS}\, \lambda_G }
which leads to the condition $\lambda_G=-\gamma_{\Omega'}\lambda_G^T
\gamma_{\Omega'}^{-1}$. The solution is  
\eqn\massc{  \lambda_G=\left(\matrix{  A  &  0  \cr
                                              0 & -A^T \cr} \right)_{64,64}, }
with $A$ any hermitian matrix implying that the state \massb\  describes a
vector boson of the gauge group $U(32)$. 
For open strings stretched between D9 and D$9'$-branes there are massless
fermion states
\eqn\massf{  | s_1, s_2, s_3, s_4 \rangle \, \lambda_F }
with $s_a= \pm {1\over 2}$ and $\sum s_a = {\rm odd}$ from the GSO
projection. Invariance under $\Omega'$ requires 
Chan-Paton factors of the form
\eqn\massd{  \lambda_F=\left(\matrix{  0  &  B  \cr
                                              C & 0\cr} \right)_{64,64}, }
with both $B^T=-B$ and $C^T=-C$. One can show that $B$ and $C$ behave 
differently under gauge transformations leading to a left-handed Majorana-Weyl
fermion in the ${\bf 496}\oplus\o{\bf 496}$ representation of 
$U(32)$. It was shown in \rsagn\  that the spectrum is anomaly-free.
Note, the $R^6$ term in the gravitational anomaly
\eqn\anom{  496\, \hat {\rm I}_{\rm Weyl} - \hat {\rm I}_{\rm 4-form}  }
vanishes and exactly for $n=32$ the trace in the antisymmetric 
representation of $U(n)$ factorizes 
\eqn\anomb{ {\rm Tr}_{A} F^6 = (n-32)\, {\rm tr}(F^6) +15\, {\rm tr}(F^2)\,
{\rm tr}(F^4) }
so that  a generalised Green-Schwarz mechanism can take over.

\subsec{D-branes in \typepp}

As we have seen in the last subsection, the closed spectrum of \typepp\
contains the same RR ($p$+1)-forms as \typeb. Thus, one expects that all
D$p$-branes, for odd $p$, are present, as well. This can also be shown using
the boundary state approach.
In \rbergab\ the action of $\Omega$ on the boundary states  
$| {\rm D}p,\eta\ra_{NSNS,RR}$ was determined
\eqn\dbranea{\eqalign{ &\Omega | {\rm D}p,\eta\ra_{NSNS}=
                    | {\rm D}p,\eta\ra_{NSNS} ,\cr
                      &\Omega | {\rm D}p,\eta\ra_{RR}=-(-i\eta)^{7-p}
                           | {\rm D}p,\eta\ra_{RR} }}
which in the case of the tachyonic \typep\ orientifold implied that only
D1,D5 and D9-branes survived the projection. 
In our case $\Omega$ is equipped with the right moving world-sheet fermion
number operation $(-1)^{F_R}$ which has the following action on the 
boundary states
\eqn\dbraneb{\eqalign{& (-1)^{F_R}| Dp,\eta\ra_{NSNS}=-|Dp,-\eta\ra_{NSNS}, \cr
                      &(-1)^{F_R} | Dp,\eta\ra_{RR}=
                           | Dp,-\eta\ra_{RR} .}}
Combining these two actions one can show that for every odd $p$ the 
D$p$-brane
\eqn\dbranec{ {1\over 2}\left( | Dp,+1\ra_{NSNS} - | Dp,-1\ra_{NSNS} + 
              | Dp,+1\ra_{RR} +
                   (-1)^{p-1\over 2} | Dp,-1\ra_{RR} \right) }
and the corresponding anti-brane are  invariant under $\Omega'$.
In particular, \typepp\ contains a self-dual D3-brane, which we will study
a bit more  in the next section.

\newsec{Massless modes on the  D3-brane}

Since we have seen that \typepp\ contains D3-branes, due to its possible 
application to non-supersymmetric gauge theories we
investigate further the effective theory on a number of parallel such 
branes. \pano
From open strings stretched between 9-branes one gets simply the dimensional
reduction of the massless spectrum discussed in section 3.5  to four
dimensions. 
{}From open strings stretching between two 3-branes or two $3'$-branes
there are massless modes of the form \massb. For $\mu$ running over
space-time values these states provide $U(M)$ gauge vectors.
For $\mu$ running over transverse coordinates we obtain six scalars
transforming in the adjoint representation. Strings stretching between
a 3 and a $3'$-brane have massless fermionic modes of the form \massf.
The four states with $s_1=-{1\over 2}$ give left-handed fermions
transforming in the $\asymm$ plus the $\basymm$ representation.
\pano
Strings stretching between 9 and 3-branes only carry massive modes,
whereas strings between 9 and $3'$-branes 
respectively $9'$ and 3-branes, have massless modes of the form
\eqn\massft{  | s_1 \rangle \, \lambda_M   }
with $s_1= \pm {1\over 2}$.
The GSO projection leaves only one state $s_1=+{1\over 2}$ transforming
in the ${\bf (32,M)}+{\bf (\o{32},\o M)}$ 
representation of the gauge group.
However as a strange fact we observe that the CPT conjugates of these 
states are absent. As listed in Table 1, one could also make the spectrum 
CPT invariant but then the states would formally transform in half gauge
multiplets\footnote{$^1$}{A similar effect happens
for the effective theory on a D5 brane in \Type\ \rwitti, where 
naively one gets
half-hypermultiplets in the 95 open string sector. However, in that case
the gauge group was $SO(32)\times Sp(2N)$ and the matter in the 95 sector
transformed in the $({\bf 32},{\bf 2N})$ representation of the gauge group. 
This representation is pseudo-real and thus there is no inconsistency.}.
It is interesting to note that the 1-loop $\beta$-function
for the $U(M)$ theory is independent of $M$.
\meno
\cl{\vbox{
\hbox{\vbox{\offinterlineskip
\def\tablespace{height2pt&\omit&&\omit&&
 \omit&\cr}
\def\tablerule{\tablespace\noalign{\hrule}\tablespace}

\hrule\halign{&\vrule#&\strut\hskip0.2cm\hfil#\hfill\hskip0.2cm\cr
\tablespace
& sector && fields &&  $U(32)\times U(M)$       &\cr
\tablerule
& 99, $9'9'$  && vectors && ({\bf Adj},{\bf 1}) & \cr 
\tablespace
&      &&  scalars && $6\times \, $  ({\bf Adj},{\bf 1})  &\cr 
\tablespace
& $99'$, $9'9$  && L-fermions &&  $4\times \{ ({\bf 496},{\bf 1}) + 
        ({\bf \o{496}},{\bf 1}) \}$&\cr
\tablerule
& 33, $3'3'$  && vectors && ({\bf 1},{\bf Adj}) & \cr 
\tablespace
&      &&  scalars && $6\times \, $  ({\bf 1},{\bf Adj})  &\cr 
\tablespace
& $33'$, $3'3$  && L-fermions &&  $4\times \{ ({\bf 1},\asymm\,) + 
        ({\bf 1},\basymm\,) \} $&\cr
\tablerule
& $93'$, $39'$ && L-fermions  && ${1\over 2}\times \{({\bf 32}, \funda )
+({\bf \o{32}},\o{\funda} )\}$ &\cr
\tablespace
& $9'3$, $3'9$ &&  &&  &\cr
\tablespace}\hrule}}}}
\cl{
\hbox{{\bf Table 1:}{\it ~~ Effective theory field content }}}
\meno 
\pano
We think that the inconsistency in the fermionic
spectrum simply reflects the fact that due to the cylinder 
amplitude for  D$9,9'$- and D$3,3'$-branes
(open strings stretching between branes of different dimensionality)
\eqn\incylnt{ A_{93}=
       {V_4\over (8\pi^2 \a')^2} \, \int_0^\infty {dt\over t^3}\, 
           2M N\,{f_3^2(e^{-\pi t}) \, f_2^6(e^{-\pi t}) -
               f_2^2(e^{-\pi t}) \, f_3^6(e^{-\pi t})
                \over {f_1^2(e^{-\pi t}) \, f_4^6(e^{-\pi t}) }}} 
there exists a repelling force between these two branes. Thus a D9- and
a D3-brane are not allowed to form neither a static configuration nor a 
bound state.
\pano
Note that $A_{93}$
contributes to the NS-NS dilaton tadpole.
This observation opens at least in principle the possibility that
the dilaton tadpole from the 9,9'-branes is cancelled by the
D3-branes. In fact, combining the 
$l \to \infty$ limit of $A_{93}$ and \tadb\ we find
that the NS-NS tadpole is cancelled provided that
\eqn\tcan{8N \, {V_6\over (8\pi^2 \a')^3} = M  ,}
where we have formally used $V_{10} = V_4 V_6$. Since the space
transverse to the $3,3'$-branes is non-compact, we conclude that
the dilaton tadpole only cancels in the limit $M \to \infty$ where the density
of D3-branes $M/V_6$ is constant.
\pano
For completeness we also present the M\"obius strip amplitude for the
$3,3'$-brane system
\eqn\moebt{ M_3 = {V_4\over (8\pi^2 \a')^2} \, 
                   \int_0^\infty {dt\over t^3}\, 2M 
   { f_2^8(i e^{-\pi t}) \over 2f_1^8(i e^{-\pi t}) } .}
This amplitude is divergence-free both in the limits $t \to 0$
and $t \to \infty$.

\newsec{Compactifications of \typepp\ to six dimensions}

Non-tachyonic compactifications of \typepp\ were considered 
in the abstract conformal field theory setting in \rangel.
In the present  and the following section we consider orientifolds
$T^{10-d}/\{G_1,\Omega'G_2\}$, $d=6,4$, with $G_1, G_2$ internal Abelian 
symmetries. We will provide a microscopic 
description of these models in terms of open strings ending on D-branes. 
The computational technique is very similar to the one presented in 
section 3, so that we restrict ourselves to the main issues. 
We still want to get models without tachyons, which highly restricts the 
number of possible internal symmetries. \pano 
In the supersymmetric case there are two possible kinds of orientifolds 
of a $\ZZ_{n}$ orbifold, which
involve different actions of world-sheet parity on the twisted sector
ground states. In the mostly studied kind one actually divides out
by $\Omega\, J'$ where $J'$ exchanges the $g$ twisted sector with the
$g^{-1}$ twisted sector. In six dimensions this leads to models with additional
tensor multiplets, cancelled tadpoles and anomaly-free perturbative
massless spectra. 
In generalising such orientifolds to our non-supersymmetric case one realizes
that generically there appear tachyons in twisted sectors, 
which are mapped under $\Omega' J'$ to tachyons in different twisted sectors. 
Thus, linear combinations
of twisted sector tachyons survive the projection. As was already observed in 
\rangel, the only  $\Omega'\, J'$ orientifolds
in six dimensions where this does not happen must have only 
twisted sectors of order two. This leaves the two possibilities
\eqn\compa{\eqalign{ &\ZZ_2:\ \{ (1 +R) \times (1 + \Omega' J')\} ,\cr
            &\ZZ_{4B}:\ \{ (1 + R) \times (1+ \Omega'J' \omega) \} , }}
where $\omega^2=R$ and $R$ is the reflection $z_i\to -z_i$
of both internal coordinates. \pano
However as pointed out in \rkaku, one can also divide out just by $\Omega$
without exchanging twisted sectors. Unfortunately, this introduces new tadpoles
from the now non-zero twisted sector Klein-bottle contributions for which we 
do  not know how to cancel them. Consistently, one does not get an anomaly
free perturbative massless spectrum. 
In the supersymmetric case \Type-heterotic duality was the guiding
principle  in determining 
what extra non-perturbative states from the \Type\ side are massless, as
well. In the \typepp\ case these pure $\Omega'$ orientifolds turn out
to be important, for in these models apparently the twisted sector
tachyons are divided out. Of course, one faces the problem of additional
non-perturbative massless states which luckily in six-dimensions might
be detected by imposing anomaly freedom. We will discuss one such example 
namely the  $\ZZ_3$ orientifold in subsection 5.3.

\subsec{The $T^4/\ZZ_2$ orientifold}

Computing the Klein bottle amplitude for this model shows that one
gets  a tadpole for both untwisted RR 10-form and RR 6-form exchange, whereas 
similar to \typepp\ in ten dimensions there is no NS-NS tadpole. 
Thus one is led to introduce  D9 and D5-branes and of course 
the corresponding D$9'$ and D$5'$-branes. The computation of the cylinder and
M\"obius amplitude
is straightforward and very similar to the computation done in \rgimpol.
\pano
One finally ends up with three RR tadpoles to be cancelled. Analogous 
to the Gimon-Polchinski model \rgimpol\ the twisted 
RR 6-form tadpoles are cancelled 
by choosing $\gamma_{R}$ traceless for both 9 and 5-branes. The remaining
two untwisted tadpole conditions are
\eqn\compb{\eqalign{  &N_9^2 - 32\, {\rm Tr}(\gamma_{\Omega',9}^T 
         \gamma_{\Omega',9}^{-1}) + 32^2 =0,\cr
         & N_5^2 - 32\, {\rm Tr}(\gamma_{\Omega' R,5}^T 
         \gamma_{\Omega' R,5}^{-1}) + 32^2 =0 .}}
Similar to the \typepp\ model there arises an uncancelled dilaton tadpole in 
the annulus amplitude. The tadpole cancellation \compb\ implies that
the two matrices $\gamma_{\Omega',9}$ and $\gamma_{\Omega' R,5}$
are symmetric. A consistent choice is
\eqn\compc{\eqalign{  &\gamma_{\Omega',9}=\left(\matrix{  0  &  1  \cr
                                          1  & 0 \cr} \right)_{2N_9,2N_9} ,
                          \quad\quad\ \ 
                     \gamma_{R,9}=\left(\matrix{  M  &  0  \cr
                                          0  & M \cr} \right)_{2N_9,2N_9},\cr
                     &\gamma_{\Omega',5}=\left(\matrix{  0  &  M  \cr
                                          M  & 0 \cr} \right)_{2N_5,2N_5} ,
                          \quad\quad
                     \gamma_{R,5}=\left(\matrix{  M  &  0  \cr
                                  0 & M \cr } \right)_{2N_5, 2N_5} .}}  
The traceless matrix $M$ is given by
\eqn\compd{ M=\left(\matrix{  1  & 0   \cr
                            0  & -1 \cr} \right)_{N_{9,5},N_{9,5}} .}  
The tadpole cancellation condition \compb\ implies that the number
of 9 and 5-branes must be 32. Next we compute the massless spectrum.
The pure bosonic massless spectrum from the closed string sector is 
presented in Table 2, where we classified the states according to
their Lorentz group $SO(4)=SU(2)\times SU(2)$ representation.
\meno
\cl{\vbox{
\hbox{\vbox{\offinterlineskip
\def\tablespace{height2pt&\omit&&
 \omit&\cr}
\def\tablerule{\tablespace\noalign{\hrule}\tablespace}

\hrule\halign{&\vrule#&\strut\hskip0.2cm\hfil#\hfill\hskip0.2cm\cr
\tablespace
& sector && spin $SU(2)\times SU(2)$      &\cr
\tablerule
& untwisted NS-NS && $(3,3)+11\times (1,1)$ &\cr
\tablespace
& untwisted R-R && $4\times (3,1)+4\times (1,3) +8\times (1,1)$  &\cr
\tablerule
& twisted NS-NS && $64\times (1,1)$ &\cr
\tablespace
& twisted R-R && $16\times (3,1) + 16\times (1,1)$ &\cr
\tablespace}\hrule}}}}
\cl{
\hbox{{\bf Table 2:}{\it ~~ Closed string spectrum of $T^4/\ZZ_2$ }}}
\meno
Thus at the massless level one gets the graviton, the dilaton, 20 self dual
2-forms, 4 anti-self dual 2-forms and 98 further scalars.
This spectrum is anomalous and must be extended by massless states from the
open string sector. In Table 3 we list the spectrum with maximally enhanced
gauge symmetry, where we have stuck together  all D5 branes on the same 
fixed point
of $R$ and where we have not turned on any  Wilson lines in the D9 brane 
gauge group.
\meno
\cl{\vbox{
\hbox{\vbox{\offinterlineskip
\def\tablespace{height2pt&\omit&&\omit&&
 \omit&\cr}
\def\tablerule{\tablespace\noalign{\hrule}\tablespace}

\hrule\halign{&\vrule#&\strut\hskip0.2cm\hfil#\hfill\hskip0.2cm\cr
\tablespace
& sector && spin && gauge    $U(16)\times U(16)\times U(16)
\times  U(16)$  &\cr
\tablerule
& 99, 55  && (2,2) && {\bf adjoint} &\cr
\tablespace
& $9'9'$, $5'5'$ && (1,1) && $4\times\{ ({\bf 16},{\bf\o{16}};
    {\bf 1},{\bf 1}) + 
   ({\bf\o{16}}, {\bf{16}};{\bf 1},{\bf 1}) + 
   ({\bf 1},{\bf 1};{\bf 16},{\bf\o{16}}) + 
 ({\bf 1},{\bf 1};{\bf\o{16}},{\bf 16}) \}$ &\cr 
\tablerule
& 95, $9'5'$  && (1,1) &&  $2\times\{ ({\bf 16},{\bf 1};{\bf\o{16}},{\bf 1}) + 
({\bf\o{16}},{\bf 1};{\bf {16}},{\bf 1}) +
 ({\bf 1},{\bf 16};{\bf 1},{\bf\o{16}}) +
 ({\bf 1},{\bf\o{16}};{\bf 1},{\bf 16}) \} $&\cr
\tablerule
& $99'$, $55'$ && (1,2) && 
  $2\times\{ ({\bf 120}\oplus {\bf\o{120}},{\bf 1};{\bf 1},{\bf 1}) + 
        ({\bf 1},{\bf 120}\oplus {\bf\o{120}};{\bf 1},{\bf 1}) + $ &\cr
& && && 
  $({\bf 1},{\bf 1};{\bf 120}\oplus {\bf\o{120}},{\bf 1})+
    ({\bf 1},{\bf 1};{\bf 1},{\bf 120}\oplus {\bf\o{120}}) \}$ &\cr
\tablespace
& && (2,1) && $2\times\{ ({\bf 16},{\bf 16};{\bf 1},{\bf 1}) + 
({\bf\o{16}},{\bf\o{16}};{\bf 1},{\bf 1}) + 
 ({\bf 1},{\bf 1};{\bf 16},{\bf 16}) + 
({\bf 1},{\bf 1};{\bf\o{16}},{\bf\o{16}}) \}$ &\cr 
\tablerule
& $95'$, $59'$ && (1,2) && $\{ ({\bf 16},{\bf 1};{\bf 16},{\bf 1}) + 
         ({\bf\o{16}},{\bf 1};{\bf\o{16}},{\bf 1}) +
            ({\bf 1},{\bf 16};{\bf 1},{\bf 16}) + 
        ({\bf 1},{\bf\o{16}};{\bf 1},{\bf\o{16}}) \} $ &\cr
\tablespace}\hrule}}}}
\cl{
\hbox{{\bf Table 3:}{\it ~~ Open string spectrum of $T^4/\ZZ_2$ }}}
\meno 
For the spectrum in Table 2 and Table 3 both the $R^4$ and the $F^4$ anomaly 
cancels \footnote{$^1$}{
Note that our spectrum does not agree with the spectrum computed in 
\rangel\ where instead of the antisymmetric representation ${\bf 120}$ there
appeared the symmetric representation ${\bf 136}$. It was explained
to us by A. Sagnotti, that both models are consistent. The freedom appears
in the definition of the $P$ modular matrix connecting the 
loop and tree channel M\"obius amplitude \rsagb.}.
There are two ways to reduce this spectrum. One way is by moving some of
the D5-branes away from the fixed point or turning on some Wilson lines
in the D9-brane gauge symmetry, respectively. 
The other way is by turning on a discrete 
background NS-NS two-form flux. Using the methods developed in
\rbackb\ we have computed some of the
resulting spectra that are presented in the appendix.  

\subsec{The $T^4/\ZZ_{4B}$ orientifold}

This is analogous to the \typeb\ orientifold discussed in \rgimjo.
Just as in that case, the change $\Omega' \to \Omega' \, \omega$
implies that the Klein bottle amplitude by itself is divergence free.
Hence, there are no branes and all matter comes from closed string sectors.
The resulting massless spectrum is shown in Table 4.
\meno
\cl{\vbox{
\hbox{\vbox{\offinterlineskip
\def\tablespace{height2pt&\omit&&
 \omit&\cr}
\def\tablerule{\tablespace\noalign{\hrule}\tablespace}

\hrule\halign{&\vrule#&\strut\hskip0.2cm\hfil#\hfill\hskip0.2cm\cr
\tablespace
& sector && $SU(2)\times SU(2)$      &\cr
\tablerule
& untwisted NS-NS && $(3,3)+7\times (1,1)$ &\cr
\tablespace
& untwisted R-R && $2\times (3,1)+6\times (1,3) +8\times (1,1)$  &\cr
\tablerule
& twisted NS-NS && $72\times (1,1)$ &\cr
\tablespace
& twisted R-R && $10\times (3,1) + 6\times (1,3) + 16\times (1,1)$ &\cr
\tablespace}\hrule}}}}
\cl{
\hbox{{\bf Table 4:}{\it ~~ Spectrum of $T^4/\ZZ_{4B}$ }}}
\meno
This matter content is anomaly-free since there are no fermions and the number
of self-dual and anti-self dual tensors is the same.

\subsec{The $T^4/\ZZ_{3}$ orientifold}

In this section we will discuss the $Z_3$ orientifold where $\Omega'$ does
not exchange the two twisted sectors. It is straightforward to compute
the massless spectrum in the closed string sector. The result is
presented in Table 5
\meno
\cl{\vbox{
\hbox{\vbox{\offinterlineskip
\def\tablespace{height2pt&\omit&&
 \omit&\cr}
\def\tablerule{\tablespace\noalign{\hrule}\tablespace}

\hrule\halign{&\vrule#&\strut\hskip0.2cm\hfil#\hfill\hskip0.2cm\cr
\tablespace
& sector && spin $SU(2)\times SU(2)$      &\cr
\tablerule
& untwisted NS-NS && $(3,3)+5\times (1,1)$ &\cr
\tablespace
& untwisted R-R && $4\times (3,1)+2\times (1,3) +6\times (1,1)$  &\cr
\tablerule
& twisted NS-NS && $54\times (1,1)$ &\cr
\tablespace
& twisted R-R && $18\times (3,1) + 18\times (1,1)$ &\cr
\tablespace}\hrule}}}}
\cl{
\hbox{{\bf Table 5:}{\it ~~ Closed string spectrum of $T^4/\ZZ_3$ }}}
\meno
Thus at the massless level one gets the graviton, the dilaton, 22 self dual
2-forms, 2 anti-self dual 2-forms and 82 further scalars.
This spectrum is anomalous and must be extended by massless states from the
open string sector and as we will see from the non-perturbative sector. 
To cancel the tadpole arising in the untwisted Klein-bottle amplitude, one has 
to introduce the same open string sector as in the $\Omega\, J'$ orientifolds. 
Thus, there are 32 D9-and 32 D$9'$-branes with the following 
action of the symmetries on the CP-factors
\eqn\purea{\eqalign{  &\gamma_{\Omega',9}=\left(\matrix{  0  & \Xi   \cr
                                          \Xi  & 0 \cr} \right)_{64,64} ,
                          \quad\quad\ \ 
                     \gamma_{\theta,9}=\left(\matrix{  \Theta  &  0  \cr
                                          0  & \Theta \cr} \right)_{64,64},\cr}}
with 
\eqn\purea{\eqalign{  
                     &\Xi=\left(\matrix{  0  &  I_8 & 0  \cr
                                        I_8 & 0 & 0 \cr
                                        0 & 0 & I_{16}\cr} \right)_{32,32} ,
                          \quad\quad
          \Theta=\left(\matrix{  e^{2\pi i\over 3}\,I_8 & 0  &  0  \cr
                                 0 & e^{-2\pi i\over 3}\,I_8 & 0 \cr
                                  0 & 0 & I_{16} \cr } \right)_{32, 32} .}}
The computation of the open string spectrum yields
\meno
\cl{\vbox{
\hbox{\vbox{\offinterlineskip
\def\tablespace{height2pt&\omit&&\omit&&
 \omit&\cr}
\def\tablerule{\tablespace\noalign{\hrule}\tablespace}

\hrule\halign{&\vrule#&\strut\hskip0.2cm\hfil#\hfill\hskip0.2cm\cr
\tablespace
& sector && spin && gauge  $U(8)\times U(8)\times U(16)$    &\cr
\tablerule
& 99, 9'9' && (2,2) && {\bf adjoint}  &\cr
\tablespace
& && (1,1) && $2\times\{ ({\bf 8},{\bf\o{8}};
    {\bf 1}) + ({\bf\o{8}}, {\bf{8}};{\bf 1}) + 
    ({\bf 8},{\bf 1}; {\bf\o{16}}) + 
    ({\bf\o{8}}, {\bf{1}};{\bf 16})+
    ({\bf 1},{\bf 8}; {\bf\o{16}}) + 
    ({\bf 1},{\bf\o{8}};{\bf 16}) \}$ &\cr
\tablerule
& $99'$  && (1,2) && 
  $2\times\{ ({\bf 8},{\bf{8}};
    {\bf 1}) + ({\bf\o{8}}, {\bf\o{8}};{\bf 1}) + 
    ({\bf 1},{\bf 1}; {\bf 120}\oplus {\bf\o{120}}) \}$ &\cr 
& && (2,1) &&  $\{ ({\bf 8},{\bf 1}; {\bf{16}})+ 
    ({\bf\o{8}}, {\bf{1}};{\bf\o{16}})+
    ({\bf 1},{\bf 8}; {\bf 16}) + 
    ({\bf 1},{\bf\o{8}};{\bf\o{16}}) \}$ &\cr
& && && $\{ ({\bf 28}\oplus {\bf\o{28}},{\bf{1}}; {\bf 1}) + 
                 ({\bf 1},{\bf 28}\oplus {\bf\o{28}};{\bf 1}) \}$ &\cr
\tablespace}\hrule}}}}
\cl{
\hbox{{\bf Table 6:}{\it ~~ Open string spectrum of $T^4/\ZZ_3$ }}}
\meno 
It is encouraging that the $F^4$ gauge anomalies do all cancel. However,
the $R^4$ anomaly is non-zero meaning that the open string spectrum can 
not be the complete spectrum. Unfortunately, since we do not know of any dual
description we might use to determine the missing states, we are restricted
to make a guess guided by anomaly cancellation and the analogy to the
\Type\ model discussed in \rkaku. 
The simplest guess for additional fermionic matter is
\eqn\purec{   9\times \{ ({\bf 28}\oplus {\bf\o{28}},{\bf{1}}; {\bf 1}) + 
                 ({\bf 1},{\bf 28}\oplus {\bf\o{28}}; {\bf 1}) \}_{(1,2)} .}
Probably, there will also be further  bosonic non-perturbative states but we 
do not know how to detect them.

\newsec{Compactifications of \typepp\ to four dimensions}

We now consider orientifolds of \typeo\ on $T^6/\{G_1,\Omega'G_2\}$
with Abelian $G_1, G_2$. To be more specific, acting on internal
complex coordinates $z_i$, $i=1,2,3$, a $G_1$ or $G_2$ element $g$
acts as $g z_i = e^{2i\pi v_i} z_i$. Absence of tachyons in closed twisted
sectors limit the allowed $\vec{v}$'s that are also constrained by
crystallographic action on the torus lattice. However, now the 
condition $ \pm v_1 \pm v_2 \pm v_3=0$ can be relaxed because our
models are non-supersymmetric from the start.\pano
One allowed model is the $\ZZ_3$ orientifold with
\eqn\ztres{\ZZ_3:\ \{(1 + \theta + \theta^2) \times 
(1 + \Omega' J')\} , }
where the generator $\theta$ has 
$\vec {v} = ({1\over 3}, {1\over 3}, - {2\over 3})$.
This example is studied in more detail in the next section.
There is also a T-dual version of the above, namely
\eqn\zseis{\ZZ_{6B}:\ \{(1 + \theta + \theta^2) \times 
(1 + \Omega' J' \beta^3) \} , }
where $\beta^2=\theta$. Another possibility is
a $\ZZ_2$ model with $G_1=G_2$ generator having 
$\vec {v} = ({1\over 2}, {1\over 2}, {1\over 2})$.

\subsec{The $T^6/\ZZ_3$ orientifold}

Computing the Klein bottle amplitude we find that there are only tadpoles
from the untwisted RR 10-form and from twisted RR 4-forms. There are only
9 and $9'$-branes and tadpole cancellation again requires $N_9 = N_{9'}=32$.
Twisted tadpoles cancel provided that
\eqn\tdrei{{\rm Tr}\, \gamma_{\theta,9} = 
{\rm Tr}\, \gamma_{\theta,9'} = -4 \   . }
This is satisfied by
\eqn\gdrei{\eqalign{ 
& \gamma_{\theta,9} = {\rm diag}\, 
(\a {\rm \bf I}_{12}, \a^2 {\rm \bf I}_{12} , {\rm \bf I}_8)  ,\cr  
& \gamma_{\theta,9'} = {\rm diag}\, 
(\a^2 {\rm \bf I}_{12}, \a {\rm \bf I}_{12} , {\rm \bf I}_8)  , }}
where $\a = e^{2i\pi/3}$. As we will see, necessarily 
$\gamma_{\theta,9'} = \gamma^*_{\theta,9}$.
In order to study the transformation of open string states it is
convenient to define the full $\theta$ embedding
\eqn\gthdrei{ \gamma_\theta = \left(\matrix{ \gamma_{\theta,9}   & 0   \cr
                            0  & \gamma_{\theta,9'} \cr}
\right)_{64,64} .}  
For the $\Omega'$ embedding we take
\eqn\opri{ \gamma_{\Omega' J'} =\left(\matrix{  0  & 1   \cr
                            1  & 0 \cr} \right)_{64,64} .}  
This is consistent with tadpole cancellation and group multiplication law. 
\pano
Let us now describe the massless spectrum. Open strings stretching between
two D9-branes or two D$9'$-branes give gauge vector states of the form
\massb, with $\mu$ a four-dimensional index. The gauge Chan-Paton factor  
must be of the form \massc\ as required by invariance under $\Omega' J'$. 
Invariance under $\theta$ also requires
\eqn\thact{\lambda_G =  \gamma_\theta \lambda_G \gamma^{-1}_\theta \ ,}  
which implies $A = \gamma_{\theta,9} A \gamma^{-1}_{\theta,9}$ and
necessarily $\gamma_{\theta,9'} = \gamma^*_{\theta,9}$. Given 
$\gamma_{\theta,9}$ in \gdrei\ we then find that the gauge group is
$U(12)\times U(12)\times U(8)$. From 99 and $9'9'$ open strings there
are also scalar states 
\eqn\scals{  \psi^i_{-{1\over 2}}|0\ra_{NS}\, \lambda_S ,}
where $i=1,2,3$ refers to the internal complex coordinates.
Invariance under $\Omega' J'$ gives
\eqn\cpscalar{ \lambda_S = \left(\matrix{  D  & 0   \cr
                            0  & -D^T \cr} \right)_{64,64} .}  
Since the world-sheet piece picks up a phase $\a$ under $\theta$,
$\lambda_S$ must satisfy
\eqn\thscalar{\lambda_S =  \a \gamma_\theta \lambda_S \gamma^{-1}_\theta \ .}  
Thus, $D = \a \, \gamma_{\theta,9} D \gamma^{-1}_{\theta,9}$. This 
determines the representations. We thus find
scalar states with multiplicities and representations given by
\eqn\screps{ 3 \times \{ ({\bf 12}, {\bf \o{12}}, {\bf 1}) + 
({\bf 1}, {\bf 12}, {\bf \o{8}}) +
({\bf \o{12}}, {\bf 1}, {\bf 8})  \} .}
Scalar states with 
$\overline{\psi}^i_{-{1\over 2}}|0\ra_{NS}\,\overline{\lambda}_S$
transform in the complex conjugate of the above. 
\pano
Open strings stretched between D9 and D$9'$-branes carry  massless
fermions of the form \massf\ with  Chan-Paton factor $\lambda_F$
as given in \massd, where $B^T = - B$ and $C^T=-C$.
The world-sheet piece acquires a phase
$e^{2i\pi(s_1 + s_2 - 2s_3)/3}$ under $\theta$ so that $\lambda_F$
must satisfy
\eqn\thfer{\lambda_F =  e^{2i\pi(s_1 + s_2 - 2s_3)/3} \,
 \gamma_\theta \lambda_F \gamma^{-1}_\theta \ .}  
This further constrains the matrices $B,C$ and gives specific
group representations for the states. We thus find that states with
$s_1= -{1\over 2}$ are massless left-handed fermions in the 
following representations and multiplicities
\eqn\ferreps{\eqalign{   
& \{ ({\bf 12}, {\bf 12}, {\bf 1}) + ({\bf 1}, {\bf 1}, {\bf 28}) + 
({\bf\o{12}}, {\bf\o{12}}, {\bf 1} ) + ({\bf 1}, {\bf 1}, {\bf\o{28}}) \} + \cr
3 \times & \{ ({\bf 66},{\bf 1},{\bf 1}) + ({\bf 1},{\bf 12},{\bf 8}) + 
({\bf 1}, {\bf\o{66}}, {\bf 1}) + 
({\bf\o{12}}, {\bf 1}, {\bf\o{8}}) \} .}}
States with $s_1= +{1\over 2}$ are the CPT conjugates of the
above. 
The open string spectrum is summarised in Table 5
\meno
\cl{\vbox{
\hbox{\vbox{\offinterlineskip
\def\tablespace{height2pt&\omit&&\omit&&
 \omit&\cr}
\def\tablerule{\tablespace\noalign{\hrule}\tablespace}

\hrule\halign{&\vrule#&\strut\hskip0.2cm\hfil#\hfill\hskip0.2cm\cr
\tablespace
& sector && spin && gauge   $U(12)\times U(12)\times U(8)$    &\cr
\tablerule
& 99, $9'9'$ && vector && {\bf adjoint} &\cr
\tablespace
&      &&  scalar && $3\times\{ ({\bf 12}, {\bf \o{12}}, {\bf 1}) + 
({\bf 1}, {\bf 12}, {\bf \o{8}}) +
({\bf \o{12}}, {\bf 1}, {\bf 8}) + c.c.\}$ &\cr
\tablerule
& $99'$ && fermion$_L$ && $\{ ({\bf 12}, {\bf 12}, {\bf 1}) + ({\bf 1}, 
{\bf 1}, {\bf 28}) + 
({\bf\o{12}}, {\bf\o{12}}, {\bf 1} ) + ({\bf 1}, {\bf 1}, {\bf\o{28}}) \}$ &\cr
\tablespace
& && && $3 \times \{ ({\bf 66},{\bf 1},{\bf 1}) + ({\bf 1},{\bf 12},{\bf 8}) + 
({\bf 1}, {\bf\o{66}}, {\bf 1}) + 
({\bf\o{12}}, {\bf 1}, {\bf\o{8}}) \} $ &\cr
\tablespace}\hrule}}}}
\cl{
\hbox{{\bf Table 7:}{\it ~~ Open string spectrum of $T^6/\ZZ_3$ }}}
\meno
This spectrum is chiral and free of non-Abelian gauge anomalies and agrees 
with the results in \rangel. Concerning the $U(1)$ factors, there is one
non-anomalous and two anomalous combinations whose anomaly could presumably
be cancelled by a generalised Green-Schwarz mechanism.

The closed string massless spectrum is computed in the standard manner.
In the untwisted sector the NS-NS states are the graviton plus 10 scalars,
including the dilaton and internal metric moduli, whereas the R-R states
are 20 scalars and one vector that arises from the 4-form. In the twisted
sector the potentially tachyonic states are instead massive but there
are 27 NS-NS massless scalars and 54 R-R massless scalars.

It is also possible to include discrete or continuous Wilson lines
to obtain new models. As explained in \rafiv, the Wilson lines
must satisfy tadpole cancellation conditions and they imply further
projections on allowed states. In the $\ZZ_3$ orientifold at hand we can
for example introduce a discrete Wilson line that gives a model with
gauge group $U(4)^8$.

\newsec{Conclusions}

In this paper we have investigated a constrained set of orientifolds of 
\typeo\ in which all tachyons are projected out. 
In detail we have discussed the ten-dimensional case, determined the D-brane
contents of this model and computed the massless spectrum. All tachyon-free
orientifolds have in common that they have unitary gauge groups only,
admit  massless fermionic matter
and that there remains an uncancelled dilaton tadpole. 
We have studied the effective gauge theory on parallel D3-branes with
the negative result that there appears an inconsistent spectrum reflecting that
D9 and D3-branes repel each other.
We briefly discussed the possibility that the dilaton tadpole could be
cancelled by the D3-branes.
\pano
Moreover, we have considered various tachyon-free
compactifications to six and four 
space-time dimensions and derived the massless anomaly-free spectra.

\bigskip\bigskip\centerline{{\bf Acknowledgements}}\pano
A.F. thanks CDCH-UCV for a research grant 03.173.98,
as well as the Quantum Field Theory
group at Humboldt-Universit\"at for financial support and very kind 
hospitality. R.B. thanks the CERN Theory division for hospitality. 

\vfill\eject
\noindent
\meno
{\bf Appendix}
\bigno\meno
i.)
\bigno
Table A1 and Table A2 list the massless closed and open string spectra of
the $T^4/\ZZ_2$ orientifold with a rank two background NS-NS 2-form 
field turned on. In the open string sector the rank of the gauge group is 
reduced by a factor of two, whereas the sector of strings stretched between
a D9 and a D5 brane is doubled. 
\meno
\cl{\vbox{
\hbox{\vbox{\offinterlineskip
\def\tablespace{height2pt&\omit&&
 \omit&\cr}
\def\tablerule{\tablespace\noalign{\hrule}\tablespace}

\hrule\halign{&\vrule#&\strut\hskip0.2cm\hfil#\hfill\hskip0.2cm\cr
\tablespace
& sector && spin $SU(2)\times SU(2)$      &\cr
\tablerule
& untwisted NS-NS && $(3,3)+11\times (1,1)$ &\cr
\tablespace
& untwisted R-R && $4\times (3,1)+4\times (1,3) +8\times (1,1)$  &\cr
\tablerule
& twisted NS-NS && $64\times (1,1)$ &\cr
\tablespace
& twisted R-R && $12\times (3,1) + 4\times (1,3) + 16\times (1,1)$ &\cr
\tablespace}\hrule}}}}
\cl{
\hbox{{\bf Table A.1:}{\it ~~ Closed string spectrum of $T^4/\ZZ_2$ 
with rk(B)=2}}}
\bigno\smno
\cl{\vbox{
\hbox{\vbox{\offinterlineskip
\def\tablespace{height2pt&\omit&&\omit&&
 \omit&\cr}
\def\tablerule{\tablespace\noalign{\hrule}\tablespace}

\hrule\halign{&\vrule#&\strut\hskip0.2cm\hfil#\hfill\hskip0.2cm\cr
\tablespace
& sector && spin && gauge      $U(8)\times U(8)\times U(8)
\times  U(8)$ &\cr
\tablerule
& 99, 55  && (2,2) && {\bf adjoint}&\cr
\tablespace
& $9'9'$, $5'5'$ && (1,1) && $4\times\{ ({\bf 8},{\bf\o{8}};
    {\bf 1},{\bf 1}) + 
   ({\bf\o{8}}, {\bf{8}};{\bf 1},{\bf 1}) + 
   ({\bf 1},{\bf 1};{\bf 8},{\bf\o{8}}) + 
 ({\bf 1},{\bf 1};{\bf\o{8}},{\bf 8}) \}$ &\cr 
\tablerule
& 95, $9'5'$  && (1,1) &&  $4\times\{ ({\bf 8},{\bf 1};{\bf\o{8}},{\bf 1}) + 
({\bf\o{8}},{\bf 1};{\bf {8}},{\bf 1}) +
 ({\bf 1},{\bf 8};{\bf 1},{\bf\o{8}}) +
 ({\bf 1},{\bf\o{8}};{\bf 1},{\bf 8}) \} $&\cr
\tablerule
& $99'$, $55'$ && (1,2) && 
  $2\times\{ ({\bf 28}\oplus {\bf\o{28}},{\bf 1};{\bf 1},{\bf 1}) + 
        ({\bf 1},{\bf 28}\oplus {\bf\o{28}};{\bf 1},{\bf 1}) + 
     ({\bf 1},{\bf 1};{\bf 28}\oplus {\bf\o{28}},{\bf 1})+$ &\cr
& && && 
  $\quad\quad ({\bf 1},{\bf 1};{\bf 1},{\bf 28}\oplus {\bf\o{28}}) \}$ &\cr
\tablespace
& && (2,1) && $2\times\{ ({\bf 8},{\bf 8};{\bf 1},{\bf 1}) + 
({\bf\o{8}},{\bf\o{8}};{\bf 1},{\bf 1}) + 
 ({\bf 1},{\bf 1};{\bf 8},{\bf 8}) + 
({\bf 1},{\bf 1};{\bf\o{8}},{\bf\o{8}}) \}$ &\cr 
\tablerule
& $95'$, $59'$ && (1,2) && $2\times\{ ({\bf 8},{\bf 1};{\bf 8},{\bf 1}) + 
         ({\bf\o{8}},{\bf 1};{\bf\o{8}},{\bf 1}) +
            ({\bf 1},{\bf 8};{\bf 1},{\bf 8}) + 
        ({\bf 1},{\bf\o{8}};{\bf 1},{\bf\o{8}}) \} $ &\cr
\tablespace}\hrule}}}}
\cl{
\hbox{{\bf Table A.2:}{\it ~~ Open string spectrum of $T^4/\ZZ_2$ with 
rk(B)=2}}}
Both the $R^4$ and the $F^4$ anomaly cancels for this spectrum.
\meno\vfill\eject
\noindent
ii.)
\smno
Table A2 and Table A4 list the massless closed and open string spectra of
the $T^4/\ZZ_2$ orientifold with a rank four  background NS-NS 2-form 
field turned on. 
\smno
\cl{\vbox{
\hbox{\vbox{\offinterlineskip
\def\tablespace{height2pt&\omit&&
 \omit&\cr}
\def\tablerule{\tablespace\noalign{\hrule}\tablespace}

\hrule\halign{&\vrule#&\strut\hskip0.2cm\hfil#\hfill\hskip0.2cm\cr
\tablespace
& sector && spin $SU(2)\times SU(2)$      &\cr
\tablerule
& untwisted NS-NS && $(3,3)+11\times (1,1)$ &\cr
\tablespace
& untwisted R-R && $4\times (3,1)+4\times (1,3) +8\times (1,1)$  &\cr
\tablerule
& twisted NS-NS && $64\times (1,1)$ &\cr
\tablespace
& twisted R-R && $10\times (3,1) + 6\times (1,3) + 16\times (1,1)$ &\cr
\tablespace}\hrule}}}}
\cl{
\hbox{{\bf Table A.3:}{\it ~~ Closed string spectrum of $T^4/\ZZ_2$ 
with rk(B)=4}}}
\meno
\cl{\vbox{
\hbox{\vbox{\offinterlineskip
\def\tablespace{height2pt&\omit&&\omit&&
 \omit&\cr}
\def\tablerule{\tablespace\noalign{\hrule}\tablespace}

\hrule\halign{&\vrule#&\strut\hskip0.2cm\hfil#\hfill\hskip0.2cm\cr
\tablespace
& sector && spin && gauge  $U(4)\times U(4)\times U(4)
\times  U(4)$     &\cr
\tablerule
& 99, 55  && (2,2) && {\bf adjoint} &\cr
\tablespace
& $9'9'$, $5'5'$ && (1,1) && $4\times\{ ({\bf 4},{\bf\o{4}};
    {\bf 1},{\bf 1}) + 
   ({\bf\o{4}}, {\bf{4}};{\bf 1},{\bf 1}) + 
   ({\bf 1},{\bf 1};{\bf 4},{\bf\o{4}}) + 
 ({\bf 1},{\bf 1};{\bf\o{4}},{\bf 4}) \}$ &\cr 
\tablerule
& 95, $9'5'$  && (1,1) &&  $8\times\{ ({\bf 4},{\bf 1};{\bf\o{4}},{\bf 1}) + 
({\bf\o{4}},{\bf 1};{\bf {4}},{\bf 1}) +
 ({\bf 1},{\bf 4};{\bf 1},{\bf\o{4}}) +
 ({\bf 1},{\bf\o{4}};{\bf 1},{\bf 4}) \} $&\cr
\tablerule
& $99'$, $55'$ && (1,2) && 
  $2\times\{ ({\bf 6}\oplus {\bf\o{6}},{\bf 1};{\bf 1},{\bf 1}) + 
        ({\bf 1},{\bf 6}\oplus {\bf\o{6}};{\bf 1},{\bf 1}) + 
     ({\bf 1},{\bf 1};{\bf 6}\oplus {\bf\o{6}},{\bf 1})+$ &\cr
& && && 
  $\quad\quad ({\bf 1},{\bf 1};{\bf 1},{\bf 6}\oplus {\bf\o{6}}) \}$ &\cr
\tablespace
& && (2,1) && $2\times\{ ({\bf 4},{\bf 4};{\bf 1},{\bf 1}) + 
({\bf\o{4}},{\bf\o{4}};{\bf 1},{\bf 1}) + 
 ({\bf 1},{\bf 1};{\bf 4},{\bf 4}) + 
({\bf 1},{\bf 1};{\bf\o{4}},{\bf\o{4}}) \}$ &\cr 
\tablerule
& $95'$, $59'$ && (1,2) && $4\times\{ ({\bf 4},{\bf 1};{\bf 4},{\bf 1}) + 
         ({\bf\o{4}},{\bf 1};{\bf\o{4}},{\bf 1}) +
            ({\bf 1},{\bf 4};{\bf 1},{\bf 4}) + 
        ({\bf 1},{\bf\o{4}};{\bf 1},{\bf\o{4}}) \} $ &\cr
\tablespace}\hrule}}}}
\cl{
\hbox{{\bf Table A.4:}{\it ~~ Open string spectrum of $T^4/\ZZ_2$ with 
rk(B)=4}}}
\meno
iii.)
\meno
In Table A.5 we list the open string spectrum one gets when  
all 32 D5 branes are moved
away from the fixed point but 16 of them still coincide and moreover 
appropriate 9-brane Wilson lines have been turned on to make the spectrum
T-invariant.
\smno
\cl{\vbox{
\hbox{\vbox{\offinterlineskip
\def\tablespace{height2pt&\omit&&\omit&&
 \omit&\cr}
\def\tablerule{\tablespace\noalign{\hrule}\tablespace}

\hrule\halign{&\vrule#&\strut\hskip0.2cm\hfil#\hfill\hskip0.2cm\cr
\tablespace
& sector && spin && gauge   $U(16)\times U(16)$    &\cr
\tablerule
& 99, 55  && (2,2) && {\bf adjoint} &\cr
\tablespace
& $9'9'$, $5'5'$ && (1,1) && $4\times\{ ({\bf Adj},{\bf 1})+
   ({\bf 1},{\bf Adj}) \} $ &\cr 
\tablerule
& 95, $9'5'$  && (1,1) &&  $4\times\{ ({\bf 16},{\bf\o{16}}) + 
({\bf\o{16}},{\bf 16}) \} $&\cr
\tablerule
& $99'$, $55'$ && (1,2) && 
  $2\times\{ ({\bf 128}\oplus {\bf\o{128}},{\bf 1}) + 
        ({\bf 1},{\bf 128}\oplus {\bf\o{128}})  \} $ &\cr
\tablespace
& && (2,1) && $2\times\{ ({\bf 136}\oplus {\bf\o{136}},{\bf 1})+ 
({\bf 1},{\bf 136}\oplus{\bf\o{136}}) \}$ &\cr 
\tablerule
& $95'$, $59'$ && (1,2) && $2\times\{ ({\bf 16},{\bf 16})+ 
         ({\bf\o{16}},{\bf\o{16}}) \} $ &\cr
\tablespace}\hrule}}}}
\cl{
\hbox{{\bf Table A.5:}{\it ~~ Open string spectrum of $T^4/\ZZ_2$ with 
9 and 5-branes not at fixed point}}}
\meno

\vfill\eject

\listrefs
\bye